\documentclass[aps,prd,secnumarabic,amssymb, amsmath,nobibnotes,nofootinbib,11pt]{revtex4}
\usepackage{amsfonts,amsmath,hyperref,url, color}
\usepackage{bm, bbm}
\usepackage{graphicx}

\newcommand{\be}{\begin{equation}}
\newcommand{\ee}{\end{equation}}
\newcommand{\bea}{\begin{eqnarray}}
\newcommand{\eea}{\end{eqnarray}}

\usepackage{color}

\begin{document}

\title{Noncommutative fields and the short-scale structure of spacetime}
\author{Michele Arzano}
\email{michele.arzano@roma1.infn.it}
\affiliation{Dipartimento di Fisica and INFN,\\ `Sapienza' University of Rome,\\ P.le A. Moro 2, 00185 Roma, EU }
\author{Jerzy Kowalski-Glikman}
\email{jerzy.kowalski-glikman@ift.uni.wroc.pl}
\affiliation{Institute for Theoretical Physics, University of Wroc\l{}aw, Pl.\ Maxa Borna 9, Pl--50-204 Wroc\l{}aw, Poland.}

\begin{abstract}
There is a growing evidence that due to quantum gravity effects the effective spacetime dimensionality might change in the UV. In this letter we investigate this hypothesis by using quantum fields to derive the UV behaviour of the static, two point sources potential. We mimic quantum gravity effects by using non-commutative fields associated to a Lie group momentum space with a Planck mass curvature scale. We find that the static potential becomes finite in the short-distance limit. This indicates that quantum gravity effects lead to a dimensional reduction in the UV or, alternatively, that point-like sources are effectively smoothed out by the Planck scale features of the non-commutative quantum fields.
\end{abstract}

\maketitle

Recent years have witnessed a surge of interest in the possibility of probing the structure of spacetime in the regime in which quantum gravity effects should become relevant by studying the scaling behaviour of the {\it spectral dimension}. This way of characterizing the dimensionality of a given space is based on a fictitious diffusion process determined, via the heat equation, by the Laplacian associated to the specific quantum gravity model under consideration \cite{Sotiriou:2011aa}. The picture that has emerged, starting with the framework of causal dynamical triangulations \cite{Ambjorn:2005db} and subsequently in a variety of approaches to quantum gravity (see e.g. \cite{Carlip:2016qrb,Lauscher:2005qz,Horava:2009if,Benedetti:2008gu,Modesto:2008jz,Benedetti:2009ge,Calcagni:2013vsa,Calcagni:2014cza,Padmanabhan:2015vma,Carlip:2015mra}), is that of a {\it dimensional reduction} at very short, Planckian, scales.

The use of the spectral dimension to explore the short scale structure of spacetime has two drawbacks: it relies on a artificial diffusion process characterized by an unphysical time parameter and can be defined only on Euclidean spaces thus applying only to Wick-rotated versions of the models at stance. This has prompted alternative proposals in which the change of dimensionality at short scales is described in terms of more intuitive or operationally better defined notions. In \cite{Amelino-Camelia:2016sru}, for example, the authors adopted certain thermodynamic quantities to characterize the dimensionality of space-time while in \cite{Alkofer:2016utc} the change of dimensionality was studied in terms of the emission rate perceived by an accelerated detector. Along these lines a particularly intriguing observation \cite{Amelino-Camelia:2013gna} suggests that in models of deformed kinematics at the Planck scale, based on deformed, non-linear, energy-momentum dispersion relations, the running of dimensionality at small scales can be actually captured by a change of the more familiar Hausdorff dimension\footnote{The Hausdorff dimension of momentum space in \cite{Amelino-Camelia:2013gna} is given by the scaling of the volume of a ball of radius $R$, e.g. in $D$-dimensional Euclidean space $V\sim R^D$ and the Hausdorff dimension is $D$.} of {\it momentum space} in the UV. As noticed in \cite{Amelino-Camelia:2013gna} such UV behaviour can be modelled by a non-trivial {\it integration measure} in four-momentum space. This feature is shared by models of deformed relativistic kinematics based on a Lie group momentum space where the deformed integration measure is determined by the curved geometry of the Lie group manifold. In the most studied examples in the literature such deformed kinematics is related to quantum group deformations of relativistic symmetries and in a configuration space picture to a non-commutative space-time as we briefly review below.

In this letter it is our aim to further explore the short-distance structure of space-time emerging in these models by using the associated non-commutative quantum fields as a probe. We study the behaviour of the potential energy between two point-like sources subject to the interaction mediated by a real scalar field living on the non-trivial momentum space. A central tool in our analysis will be the generating functional of the free quantum scalar field coupled to the sources. %Upon introducing a two static sources we are able to analytically derive the form of the potential between them

Quite strikingly we obtain that, unlike for ordinary local quantum fields, this potential energy {\it does not} diverge in the zero-distance limit. %This result agrees with the running of the spectral dimension to $3$ found in the literature ({\bf (MA: what about other choices of the Laplacian? It would be interesting to have at least another example)}).
This indicates that Planck scale effects encoded in the non-abelian group manifold structure of momentum space introduce an effective dimensional reduction in the UV. We show that such effect does not depend on the choice of kinetic operator of the theory (as it is the case for the analysis of the spectral dimension of these models) and that the UV finiteness of the potential is associated to an effective smearing of the point sources as ``seen" by the non-commutative interaction carrier field.\\

%\section{From Planck scale kinematics to deformed QFT}
The basic feature of the model of deformed kinematics we will be considering is that momentum space, rather than being ordinary flat Minkowski space, is described by the non-abelian Lie group $AN_3$, a subgroup of the five dimensional Lorentz group $SO(4,1)$ (see \cite{Freidel:2007hk} for details). As a manifold this group is a ``half'' of  four-dimensional de Sitter space whose cosmological constant $\kappa^2$ is representative of an energy scale which can be identified with the Planck energy. The $AN_3$ group can be obtained by exponentiating the $\mathfrak{an}_3$ Lie algebra
\be\label{kMink}
[X_0, X_a] = \frac{i}{\kappa} X_a\,, \quad [X_a, X_b] = 0\,; \quad a = 1,\ldots,3
\ee
known in the literature as $\kappa$-Minkowski non-commutative space-time. Notice that in order for the generators of the Lie algebra to carry dimension of length, one has to introduce the constant $\kappa$ with dimension of energy. This sets the UV scale associated to the curvature $1/\kappa^2$ of the momentum group-manifold (in the limit $\kappa\rightarrow \infty $ the algebra of coordinates becomes abelian and one recovers the usual flat momentum space).

One of the key features of field theories defined on the $AN_3$ four-momentum space is that ordinary plane waves are replaced by group elements i.e. exponentials of the non-commuting algebra elements (\ref{kMink}).  For instance, a choice of {\it normal ordering} for the plane waves can be associated to a given parametrization of the group element \cite{Arzano:2010jw}. In particular for  ``time-to-the-right'' ordered plane waves
\be\label{Kexp}
e_k = e^{-i \vec{k} \cdot \vec{X}} e^{i k_0 X_0}\,,
\ee
the real parameters $k_0$, $\vec{k}$ are coordinates on the $AN_3$ group momentum space and are known as ``bicrossproduct" \cite{Majid:1994cy} or ``horospherical" coordinates \cite{Arzano:2014jua}.

There is yet another important coordinate system on the $AN_3$ momentum space, describing its embedding into the five-dimensional Minkowski space. Such ``embedding coordinates'' are related to the bicrossproduct coordinates by the following coordinate transformation
\begin{align}\label{eq:2.3}
p_0 &= \kappa \sinh\left(\tfrac{k_0}{\kappa}\right) + \frac{1}{2\kappa} e^{k_0/\kappa} \vec{k}^2\,, \nonumber\\
\vec{p} &= e^{k_0/\kappa} \vec{k}\,, \nonumber\\
p_4 & = \kappa \cosh\left(\tfrac{k_0}{\kappa}\right) - \frac{1}{2\kappa} e^{k_0/\kappa} \vec{k}^2\,.
\end{align}
One can easily check that the embedding coordinates above satisfy the constraints
\be\label{dScon}
-p_0^2 + \vec{p}^2 + p_4^2 = \kappa^2\,,\,\,\,\,\,\,\, p_0 + p_4 > 0\,,
\ee
which define the manifold of the $AN_3$ group as submanifold of four-dimensional de Sitter space. Notice that taking the flat limit $\kappa \rightarrow +\infty$ one has $p_0 \rightarrow k_0$, $\vec{p} \rightarrow \vec{k}$ but $p_{4} \rightarrow +\infty$ and therefore $p_{4}$ can be identified as the ``auxiliary" momentum in embedding coordinates to be considered as a function of energy $p_0$ and spatial momenta $\vec{p}$ via (\ref{dScon}).

The kinematical and relativistic properties of the $AN_3$ group valued momenta, action of Lorentz transformations and composition of momenta, are described by a Hopf algebra deformation of the Poincar\'e algebra known as $\kappa$-Poincar\'e \cite{Lukierski:1991pn}, \cite{Lukierski:1992dt}, \cite{Majid:1994cy}. For the purposes of the present work it will be sufficient to describe the action of translation generators on plane waves and the definition of the (deformed) Casimir mass invariant. Translation generators, as in the ordinary case, act on plane waves as derivatives but in this case the Leibniz rule for acting on products of plane waves will be non-linear and non-symmetric, a typical feature of symmetry generators belonging to a non-trivial Hopf algebra. On a single plane wave $e_{k} = e^{-i \vec{k}\cdot \vec{X}} e^{i k_0 X_0}$ the translation generators $P_{\mu}$ act according to
\be\label{trplane}
P_{\mu}\, e_{k} = p_{\mu}(k_0, \vec{k})\, e_{k}\,,
\ee
with eigenvalues $p_{\mu}(k_0, \vec{k})$ given by the first four entries of (\ref{eq:2.3}). This action is associated to a type of {\it non-commutative} differential calculus (the interested reader can consult \cite{Sitarz:1994rh} and  \cite{Freidel:2007hk} for full details on this choice of calculus). Let us just mention in passing that covariance requires such differential calculus to be five-dimensional with $P_{\mu}\equiv\hat{\partial}_{\mu}$ acting on plane waves as in (\ref{trplane}) and the $\hat{\partial}_{4}$, the additional derivative, as  $\hat{\partial}_{4}\, e_{k} = (1- p_4(k_0, \vec{k}))\, e_{k}$.
There is a natural d'Alembertian operator associated to this five-dimensional calculus which, in terms of the eigenvalues of the translation operators described above, realizes the invariant
\be\label{calC}
C(p)= p_0^2 -\vec{p}{\,}^2
\ee
which formally corresponds to the ordinary relativistic mass Casimir. Such invariant has an intuitive geometrical meaning in terms of sub-manifolds of de Sitter momentum space spanned by hyper-surfaces of constant auxiliary momentum $p_4$.

A free quantum scalar field defined on the Lie group $AN_3$ can be constructed in a rather straightforward way in terms of a path integral \cite{AmelinoCamelia:2001fd}. Indeed path integrals for fields defined on (several copies) of a Lie group are well known and have been widely studied in the quantum gravity literature under the name of {\it group field theories} (see \cite{Livine:2009zz} for a discussion oriented towards non-commutative models and relevant references). From this point of view the ``deformed" quantum fields we are considering here can be seen as a one-dimensional group field theory with a non-trivial kinetic term represented by the Casimir $C(p)$.

As mentioned above, in order to explore the dimensionality of space-time as probed by a non-commutative field, we will study the interaction between two point sources mediated by the exchange of a massless scalar particle. From a path integral point of view this information is encoded in the partition function of the scalar field coupled to external sources which, as in ordinary QFT, represents the vacuum-to-vacuum transition amplitude, $Z \equiv \langle 0| e^{-iHT}|0\rangle$. Such partition function can be computed by functional integration of the action for the field and sources. Our starting point will be the non-commutative partition function written in terms of the star product $\star$ associated with the choice of the non-commutative plane waves (\ref{Kexp}), introduced in \cite{Freidel:2007hk} and discussed in details in \cite{KowalskiGlikman:2009zu},
\begin{align}\label{ncpartfun}
Z[J]=\frac{1}{Z[0]}\int\!\!\mathcal{D}\phi\ \exp\left(\frac{i}{2}\int\!d^4 x\,(\partial_{\mu}\phi(x)\star \partial_{\mu}\phi(x)+\phi(x)\star J(x)+J(x)\star \phi(x))\right),
\end{align}
where $\partial_{\mu}$ are derivatives determined by the covariant calculus discussed above.  It turns out that the expression above can be recast in terms of ordinary derivatives and products in a {\it non-local} form using the relation\footnote{It should be noted that, as discussed in  \cite{KowalskiGlikman:2009zu}, this simplification is possible only in the case of bilinear expressions and does not apply to higher order polynomials in fields.}
\be
\psi \star \phi \equiv \psi \sqrt{1- \frac{\Box}{\kappa^2}} \phi + \textrm{total derivative}\,,
\ee
so that \eqref{ncpartfun} becomes
\begin{align}\label{partfun2}
Z[J]=\frac{1}{Z[0]}\int\!\!\mathcal{D}\phi\ \exp\left(\frac{i}{2}\int\!d^4 x\,(\phi\, (\Box\, \sqrt{1- \Box/\kappa^2})\, \phi+\phi\, \sqrt{1- \Box/\kappa^2}\, J+J \sqrt{1- \Box/\kappa^2} \,\phi)\right)\,.
\end{align}
As it is well known \cite{Dittrich:2012kf,Guedes:2013vi}, for fields defined on a Lie algebra spacetime one can define a non-commutative Fourier transform to functions on the Lie group momentum space. In the $\kappa$-Minkowski case such Fourier transform can be written \cite{KowalskiGlikman:2009zu} using the explicit form of the $\star$-product between function and plane wave as
\be\label{p4ft}
\tilde{\phi}(P)= \frac{p_4}{\kappa}\int d^4 x\, e^{-ipx} \phi(x)\,.
\ee
where $e^{-ipx}$ is a {\it commutative} plane wave. The inverse transform is given by
\be\label{ip4ft}
\phi(x)= \int_{AN_3} \frac{d\mu(p)}{(2\pi)^4} e^{ipx} \tilde\phi(p)
\ee
where $d\mu(p)$ is the Lorentz invariant Haar measure on the momentum Lie group $AN_3$ which, in embedding coordinates, can be written down in terms of the usual Lebesgue measure on $\mathbb{R}^{4,1}$ as
\be
d\mu(p) = d^{5} p\, \delta(-p_0^2 + \vec{p}^2 + p_4^2 - \kappa^2)\, \theta(p_0 + p_4)\,.
\ee
It should be noted that in the integral (\ref{ip4ft}) we integrate over the $AN_3$, not the $R^4$ manifold. This will have its impact on the range of integration in the integral (\ref{w12}) below, where $|\vec p|$ is forced to belong to the interval from $0$ to $\kappa$.

Under Fourier transform the non-local operator $\sqrt{1- \Box/\kappa^2}$ is mapped to $p_4$ and thus the resulting partition function can be expressed in momentum space as
\be
Z[J]=\frac{1}{Z[0]}\int\!\!\mathcal{D}\phi\ \exp \left(\frac{i}{2}\int\!\frac{d\mu(p)}{(2\pi)^4} \,\left(\phi(-p)\,\mathcal{C}(p)\,p_4 \phi(p)+ \phi(-p)\, p_4 J(p)+J(-p)\, p_4 \phi(p)\right)\right)
\ee
where we removed the tilde from the field in order to avoid cluttering the notation. The integration over the field $\phi$ can be carried out as a ordinary gaussian integral to obtain
\be\label{kpartfunction}
Z(J) = Z(0) \exp\left(-\frac{i}{2} \int \frac{d^4 p}{(2\pi)^4 } \frac{J(-p)J(p)}{p_4\,\,\mathcal{C}(p)}\right)\,,
\ee
where we assume that a small imaginary term has been added to $\mathcal{C}(p)$ to render the expression well defined. The effective action $W(J)$ defined by
\be
Z (J) /Z(0) = e^{i W(J)}
\ee
will be proportional to the energy exchanged by a configuration of sources whose profile is described by $J(p)$. As it is well known in the case of the effective action for a standard local quantum field theory, if $J(p)$ represents two static point-like sources located at distance $r$ the potential energy derived from $W(J)$ scales as $1/r$. Let us now investigate how this result changes in the case of the deformed theory.

For two static sources located at $\vec{x}_1$ and $\vec{x}_1$ one customarily chooses $J(x)= J_1(x) + J_2(x) = q_1\delta(\vec{x}-\vec{x}_1)+ q_2\delta(\vec{x}-\vec{x}_2)$, where $q_{1/2}$ are the charges of the point sources. The corresponding momentum space expressions, obtained by using the Fourier transform \eqref{p4ft}
take the form
\begin{equation}\label{sourcem}
  J_i(p) =q_i \frac{p_4}{\kappa} \delta(p_0) e^{-i\vec{p}\,\vec{x}_i}\,,\,\,\,\,\,\,\, i=1,2\,.%\quad J_2(p) =q_2 \frac{p_4}{\kappa} \delta(p_0) e^{-i\vec{p}\,\vec{x}_2}\,.
\end{equation}
We can now plug the expressions for the sources (\ref{sourcem}) in the partition function (\ref{kpartfunction}) and consider the ``interaction" terms of the effective action containing the product of $ J_1(p)$ and $J_2(p)$, which after integrating over $p_0$ take the form
\begin{equation}\label{W12}
 W_{ij}(J)
= -\frac{q_iq_j}{2} \, \delta(0) \, \int \frac{d^3 p}{(2\pi)^3}\,p_4(0,\vec{p})\, \frac{e^{i \vec{p}\cdot (\vec{x}_i - \vec{x}_j)} }{\mathcal{C}(0,\vec{p})}\,,\,\,\,\,\,\,i\neq j,\,\,\, i,j=1,2
\end{equation}
In our case $\mathcal{C}$ is just the standard Lorentz invariant (\ref{calC}) and therefore, after performing integration over angles the integral above gives
\begin{equation}\label{w12}
  W_{ij}(J) =\frac{1}{2\pi^2 r}\, \frac{q_iq_j}{2} \,T \int_0^\kappa dp\,\sqrt{1 - p^2/\kappa^2}\, \frac{\sin(p r)}{p}
\end{equation}
where $T=\delta(0)$ is the (formally infinite) time of the process.
The integral above can be analytically evaluated. Taking into account the fact that the expression for $W_{ij}$ computed above is related to the potential energy between two sources by
\begin{equation}\label{VW}
V(r) = -2W_{ij}/T
\end{equation}
%(the factor 2 results from the fact that there are two equal contributions to the potential energy, one from $W_{12}$ and another from $W_{21}$),
we obtain the potential energy
\begin{equation}\label{5}
V (r) =-  \frac{q_1q_2}{8\pi r}\,\left[  \pmb{J}_1(r \kappa ) \left(-2+\pi r \kappa\,\pmb{H}_0(r \kappa )\right)+r\kappa\, \pmb{J}_0(r \kappa ) \left(2-\pi  \pmb{H}_1(r \kappa )\right) \right],
\end{equation}
where $\pmb{J}_{\nu}$ and $\pmb{H}_{\nu}$ are Bessel $J$ and Struve $H$-functions.

The plot of the potential above for small and large values of $r$ are depicted in Fig.\ 1. It follows from the asymptotic expansion of Bessel and Struve functions that at large distances
\begin{equation}\label{5a}
  V(r) \sim -\frac{q_1q_2}{4\pi r}\left(1- \frac{\cos(\kappa r) + \sin(\kappa r)}{\sqrt{\pi}}\,\frac{1}{(\kappa r)^{3/2}} + O\left(\frac1{(\kappa r)^{5/2}}\right)\right)\,, \quad \mbox{for $r\rightarrow\infty$}
\end{equation}
We see that for large $r$ (compared to $1/\kappa$, which is of order of the Planck length, $10^{-35}$ m)
the potential $V(r)$ is the standard $1/r$ potential with the leading modification of the form $1/(\kappa r)^{3/2}$, which for the charges separation of $1$ meter, say, is of the relative size of $\sim10^{-53}$ and thus completely negligible for all practical purposes. On the other hand in the zero distance limit the potential is finite and approaches the Planckian value of $V(0) = -\frac{q_1q_2\kappa}{8 \pi}$. This is a remarkable result. To understand its relevance recall that it has been argued that the deformed kinematic associated with $\kappa$-deformations and $AN_3$ group valued momenta arises in a particular, ``flat" limit of quantum gravity \cite{AmelinoCamelia:2011bm}. Such UV finiteness of the deformed kinematics\footnote{Let us mention that in a recent work by one of the authors \cite{Arzano:2017mdp}, a somewhat related result concerning the UV finiteness of the Euclidean two-point function of a model with compact momentum space was discussed using heat kernel techniques.} can be understood as the manifestation of the long time expected property of quantum gravity acting as an universal UV regulator.

\begin{figure}
    \centering
     \includegraphics[width=0.45\textwidth]{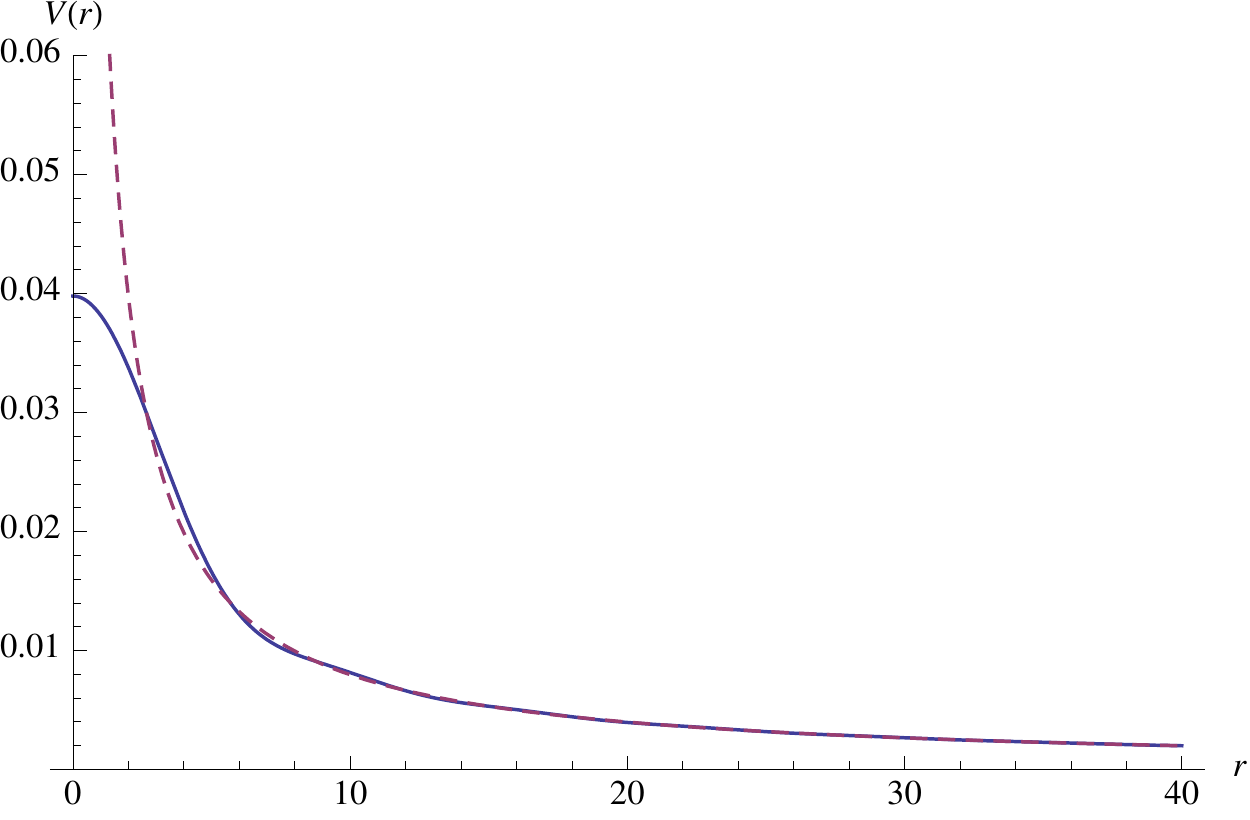}
       \qquad
    ~ %add desired spacing between images, e. g. ~, \quad, \qquad, \hfill etc.
        \includegraphics[width=0.45\textwidth]{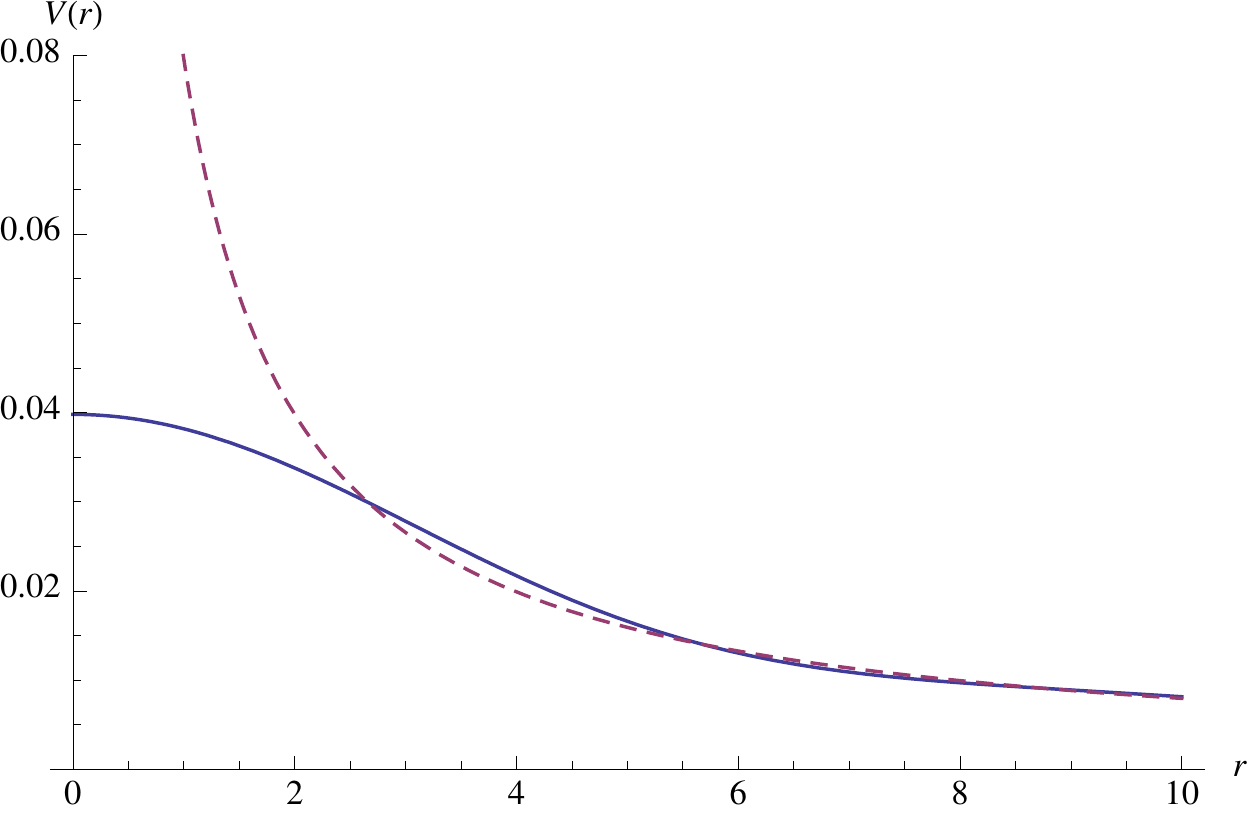}
           \caption{The potential $V(r)$ in (\ref{5}) for $\kappa=1$ and for unit charges $q_1$ and $q_2$ (solid line) vs. the ordinary $1/4\pi r$ potential (dashed line). On the right a detail of the Planckian regime.}
\end{figure}

The regular short distance behaviour of \eqref{5} can be also interpreted as an indication of an effective dimensional reduction in the UV. Indeed, in $D$ spacetime dimension the static potential scales as $V_D(r) \sim 1/r^{D-3}$ and thus the finite value fo the potential in the UV can be seen as a reduction of the dimension of spacetime to 3 at short scales. It is interesting to notice that such value agrees with the UV behaviour of the {\it spectral dimension} studied in \cite{Arzano:2014jfa} for the particular choice of momentum space Laplacian corresponding to the Wick rotated Casimir $\mathcal{C}(p)$.

It should be noted, however, that, as shown in \cite{Arzano:2014jfa}, the spectral dimension exhibits drastically different behaviours according to the choice of Laplacian. It is thus interesting to check how the expression for the potential (\ref{5}) is affected when the kinetic operator $p^2$ is replaced with a more general non-linear function of $\vec{p}$ and $p_0$. Let us notice that a general kinetic operator ${\cal C}(p_0=0,\vec{p})$ which is a regular functions of $p=|\vec{p}|$ for $p\in(0,\kappa)$, in the IR limit $p\ll\kappa$ must reduce to the usual, undeformed expression ${\cal C}(p_0=0,p)\sim p^2$. The integral (\ref{w12}) is now replaced by
\begin{equation}\label{6}
 V(r) \sim\frac1r\, \int_0^\kappa  {p\,dp}\, {\sqrt{1 - p^2/\kappa^2}}\, \frac{\sin(p r)}{{\cal C}(p_0=0,\vec{p})}\,.
\end{equation}
Since ${\cal C}(p_0=0,\vec{p})\sim p^2$ for small $p$ the integrand remains finite at $p=0$. As a result of the regularity of ${\cal C}(p)$ on the whole finite integration range $(0,\kappa)$ the integral is finite.
%The problem can arise if we have to do with the dispersion relation going to zero at $p\rightarrow\kappa$ very fast,    as $(1 - p^2/\kappa^2)^{3/2+\epsilon}$ with $\epsilon\geq0$. Such dispersion relation would be pretty weird because it would start at zero, have a maximum and then go to zero again, and we do not know of any model in which such dispersion relation would appear.
As a consequence, for a generic ${\cal C}(p_0=0,\vec{p})$ the potential $V(r)$ does not diverge at $r=0$. Too see this it is sufficient to expand $V(r)$ near $r\sim0$ and to notice that the leading order term will be a $r$-independent finite expression, with all the sub-leading terms vanishing in the limit $r\rightarrow0$. Thus the potential $V(r)$ remains finite in the UV limit again indicating an effective dimensional reduction to 3. This conclusion holds, for example, in the case of the so called `bicrossproduct basis' dispersion relation
\be\label{bcplap}
{\cal C}(k) = 4\kappa^2\sinh^2(k_0/2\kappa) - e^{k_0/\kappa}\, \vec{k}{}^2\,.
\ee
which results in
\begin{equation}\label{7}
  {\cal C}(p_0=0,p)= 2\kappa^2\left(1-\sqrt{1-\frac{p^2}{\kappa^2}}\right)\,.
\end{equation}
which is a smooth function of $p$ for $p\in (0,\kappa)$ as required. As shown in \cite{Arzano:2014jfa} the spectral dimension determined by the bicrossproduct Laplacian \eqref{bcplap} does not lead to dimensional reduction, rather in the UV becomes higher reaching the value of 6 for vanishing diffusion times. We thus see that the effective dimensionality probed by the interactions mediated by the non-commutative field {\it does not} appear to be sensitive to the choice of Laplacian and thus the type of dimensional reduction associated to a UV finite potential can be taken as a universal feature of $\kappa$-deformed models.\\

Classically we know that the usual $1/r$ potential derives from a solution of the Poisson equation with a point-like delta source. As a final exercise, in order to get additional intuition on the effects of deformations, let us determine what kind of source of the Poisson equation would be effectively associated to the deformed potential (\ref{5}). We do so by simply evaluating the Laplacian of the potential which gives the following source profile:
\begin{equation}\label{6}
\varrho(r)=  \frac1{r^2}\partial_r r^2\partial _r\frac{ {\cal V}(r)}{q_1} = \frac{q_2\kappa}{4\pi r^2} \, \pmb{J}_2(\kappa r)
\end{equation}
\begin{figure}
    \centering
        \includegraphics[width=0.4\textwidth]{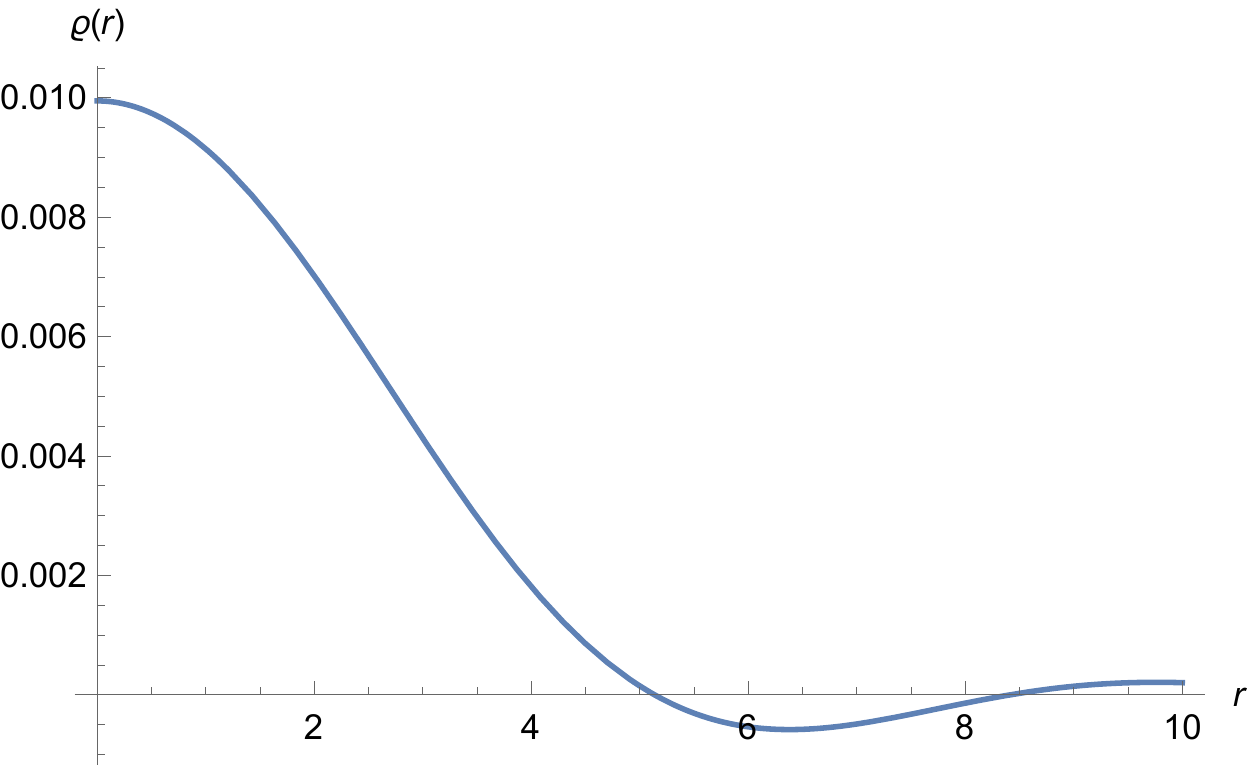}
           \caption{The source profile (\ref{6}) for $\kappa=1$ and for the unit charge $q_2$.}
\end{figure}
As it can be seen from Fig.\ 2 the source is finite at $r=0$ and then it goes to zero rapidly, with small fluctuations superimposed on the overall fall-off. This behaviour is consistent with the picture of the profile \eqref{6} as a {\it smoothed out} version of the ordinary point-like density. Indeed in the limit $\kappa\rightarrow\infty$ it is easily seen that the function \eqref{6} is very peaked at the origin and almost zero everywhere else, approaching a point-like profile while for finite $\kappa$ the source looks like an extended Planck sized lump of charge.\\

Let us remark that, in principle, the results we obtained readily generalize to the case of massless spin 1 and spin 2 carriers. Although a full fledged $\kappa$-quantum field theory with a Lie group momentum space for arbitrary spin fields is still to be formulated it is known that the representation theory of one-particle states is that same as in the case of the standard relativistic quantum fields based on the Poincar\'e group \cite{Ruegg:1994bk}. This strongly suggests that the tensor structure of the Green functions is exactly the same as in ordinary field theory and thus for the electromagnetic interactions mediated by spin-1 massless particles we expect to get the standard repulsive, while in the case of spin-2, the standard attractive force between electric charges, and masses, respectively while the space dependent part of the potential will still be given by the modified formula (\ref{5}) in both these cases.

The analysis we presented shows that $\kappa$-deformed field theories based on the $AN_3$ Lie group momentum space naturally exhibit a finite UV behaviour for the interaction mediated by a deformed field carrier. Such UV regularity does not depend on the particular choice of Laplacian, or basis of translation generators, and can be interpreted as an indication of an effective dimensional reduction of space-time at short distances. Our results confirm the intuitive view that Planck scale effects due to the presence of the deformation parameter $\kappa$ might provide a fuzzy picture of space-time showing that point-like sources do not lead to divergent interactions or, alternatively, that it is impossible to have sources localized at a point. We postpone a detailed study of such fundamental limitation in localizing sources in an arbitrarily small region of space to future studies.

\section*{Acknowledgements}
For JKG this work was supported  by funds provided by the National Science Center, projects number 2011/02/A/ST2/00294 and
2014/13/B/ST2/04043.

\end{document}